\documentclass[prc,twocolumn,nofootinbib,showpacs,floatfix,letterpaper]{revtex4-1}
\usepackage{amssymb,amsmath}
\usepackage{graphicx}
\usepackage{tikz,longtable,multirow}
\usepackage{graphics}
\usepackage{epsfig}
\usepackage{supertabular}
\usepackage{dcolumn}

\begin{document}

\title{Non-yrast spectra of odd-$A$ nuclei in a model of  coherent quadrupole-octupole motion}

\author{N. Minkov\textsuperscript{1,2}\email{nminkov@inrne.bas.bg},
S. Drenska\textsuperscript{1}, K. Drumev\textsuperscript{1}, M. Strecker\textsuperscript{2},
H. Lenske\textsuperscript{2}, W. Scheid\textsuperscript{2}}

\affiliation{\textsuperscript{1}Institute of Nuclear Research and Nuclear Energy, Bulgarian
Academy of Sciences, Tzarigrad Road 72, BG-1784 Sofia, Bulgaria \\
\textsuperscript{2}Institut f\"{u}r Theoretische Physik der Justus-Liebig-Universit\"at,
Heinrich-Buff-Ring 16, D--35392 Giessen, Germany}

\date{\today}

\begin{abstract}
The model of coherent quadrupole and octupole motion (CQOM) is applied to describe
non-yrast split parity-doublet spectra in odd-mass nuclei. The yrast levels are described as
low-energy rotation-vibration modes coupled to the ground single-particle (s.p.) state, while
the non-yrast parity-doublet structures are obtained as higher-energy rotation-vibration modes
coupled to excited s.p. states. It is shown that the extended model scheme describes both the
yrast and non-yrast quasi parity-doublet spectra and the related B(E1) and B(E2) transition
rates in different regions of heavy odd-$A$ nuclei. The involvement of the
reflection-asymmetric deformed shell model to describe the single-particle motion and the
Coriolis interaction on a deeper level is discussed.
\end{abstract}


\maketitle

\section{Introduction}
\label{intro} The observation of positive- and negative-parity states connected
by E1 and E3 transitions in atomic nuclei is usually explained with the
presence of quadrupole-octupole deformations \cite{EG87,BN96}.  In the
even-even nuclei one typically observes alternating-parity bands, whereas in
odd-mass nuclei the spectrum is characterized by a quasi parity-doublet
structure \cite{BN96}. The low-lying (yrast) structure of the
quadru\-pole-octupole spectra was relatively well studied within different
microscopic and collective model approaches \cite{Leand82}--\cite{Buck08} (see
also \cite{BN96} and references therein). However, the interpretation and the
model classification of the higher, non-yrast parts of these spectra is still
limited.

Recently the model of Coherent Quadrupole-Octupole Motion (CQOM)
\cite{b2b3mod,b2b3odd} was applied to describe non-yrast collective bands with
positive and negative parities in even-even nuclei together with attendant
B(E1), B(E2) and B(E3) transition probabilities \cite{MDSSL12}. It was shown
that couples of $\beta$-bands and higher-energy (non-yrast) negative-parity
bands can be interpreted within the model framework in the same way as the
yrast alterna\-ting-parity bands. On this basis it was concluded that the
octupole degrees of freedom have a persistent role at higher energies and the
quadrupole-octupole structure of the spectrum develops towards the non-yrast
region of collective excitations \cite{MDSSL12}.

The purpose of the present work is to examine the capability of the CQOM model
scheme to describe non-yrast quadrupole-octupole excitations in odd-mass nuclei
by extending the original consideration proposed in \cite{b2b3odd}. For this
reason it is assumed that some non-yrast level sequences with positive and
negative parities observed in these nuclei can be associated to a higher-energy
quasi parity-doublet structure of the spectrum. It is supposed that such a
structure can appear as the manifestation of higher-energy quadrupole-octupole
rotation-vibration modes coupled to the single-particle motion. The coupling of
the odd nucleon to the even-even nuclear core as well as the Coriolis
interaction are taken into account phenomenologically. At the same time the
implemented approach is supposed to pave the way for a subsequent microscopic
treatment of the odd-nucleon degrees of freedom. In this meaning the present
work may be considered a necessary step towards a deeper understanding of the
mechanism which governs the evolution of quadrupole-octupole dynamics in the
higher-energy part of the spectrum in odd-mass nuclei.

This work is organized as follows. In Sec.~\ref{sec:2} the CQOM model formalism
for the split parity-doublet bands and its features in the non-yrast part of
the spectrum are briefly presented. In Sec.~\ref{sec:3} numerical results and a
discussion on the application of the model to a number of odd-mass nuclei in
the rare-earth and actinide regions are presented. In Sec.~\ref{sec:4}
concluding remarks are given.

\section{Model of Coherent Quadrupole--Octupole Motion}
\label{sec:2}

The model Hamiltonian for odd-mass nuclei is taken as \cite{b2b3odd}
\begin{eqnarray}
H_{\mbox{\scriptsize qo}}&=&-\frac{\hbar^2}{2B_2}\frac{\partial^2}{\partial\beta_2^2}
-\frac{\hbar^2}{2B_3}\frac{\partial^2}{\partial\beta_3^2}+
U(\beta_2,\beta_3,I,K,\pi a) , \label{Hqo}
\end{eqnarray}
where $\beta_2$ and $\beta_3$ are axial quadrupole and octupole deformation variables,
respectively, and
\begin{equation}
U(\beta_2,\beta_3, I,K,\pi a)=\frac{1}{2}C_2{\beta_2}^{2}+
\frac{1}{2}C_3{\beta_3}^{2} + \frac{X(I,K,\pi a)}{d_2\beta_2^2+d_3\beta_3^2} .
\label{Ub2b3I}
\end{equation}
Here $B_2$ $(B_3)$, $C_2$ $(C_3)$ and $d_2$ ($d_3$) are quadrupole (octupole) mass,
stiffness and inertia parameters, respectively. The quantity
\begin{eqnarray}
X(I,K,\pi a)&=& \frac{1}{2}\biggl[ d_0+I(I+1)-K^2
\nonumber \\
&+&\left. \pi a \delta_{K,\frac{1}{2}}(-1)^{I+1/2}
\left(I+\frac{1}{2}\right)\right ] \ ,
\label{Xdecoup}
\end{eqnarray}
involves the total angular momentum $I$, its third projection $K$ and the decoupling factor
$a$ for the intrinsic states with $K=1/2$. The parameter $d_0$ determines the potential core at
$I=0$. In the present work the decoupling factor is considered as a model parameter and is
adjusted to the experimental data. In \cite{MDSS10} we show that a microscopically
determined effect of the Coriolis interaction can be included in CQOM through an appropriate
parity-projection particle-core coupling scheme in which the decoupling factor is calculated
by using a reflection-asymmetric deformed shell model \cite{qocsmod}.

Under the assumption of coherent quadrupole-octupole oscillations with a frequency
\begin{eqnarray}
\omega=\sqrt{\frac{C_2}{B_2}}=\sqrt{\frac{C_3}{B_3}}
\equiv \sqrt{\frac{C}{B}},
\end{eqnarray}
and after introducing ellipsoidal coordinates
\begin{eqnarray} \eta=\left[
\frac{(d_2\beta_2^2+d_3\beta_3^2)}{d}\right]^{\frac{1}{2}} \ \ \
\mbox{and}\ \ \ \phi=\arctan\left( {\frac{\beta_3}{\beta_2}
\sqrt{\frac{d_3}{d_2}}}\right )\ , \nonumber
\end{eqnarray}
such that
\begin{eqnarray}
\beta_{2}=p\eta\cos\phi, \qquad \beta_{3}=q\eta\sin\phi,
\label{polar}
\end{eqnarray}
with
\begin{eqnarray}
p=\sqrt{\frac{d}{d_{2}}},\ \ \ q=\sqrt{\frac{d}{d_{3}}},\ \ \
d=\frac{1}{2}(d_{2}+d_{3}),
\label{pqd}
\end{eqnarray}
the collective energy of the system is obtained in the following analytic form \cite{b2b3odd}
\begin{eqnarray}
E_{nk}(I,K,\pi a) &=&\hbar\omega \left[ 2n+1+\sqrt{k^2+bX (I,K,\pi a)}\right],
\label{enspect}\\
n&=&0,1,2,...\ , \ \ k=1,2,3,...\ ,
\nonumber
\end{eqnarray}
where $n$ has the meaning of a radial $(\eta)$ quantum number, $k$ corresponds to an
angular $(\phi)$ quantum number and $b=2B/(\hbar^2 d)$ is considered as a parameter. The
quadrupole-octupole vibration wave function is
\begin{eqnarray}
\Phi^{\pi_{\mbox{\scriptsize c}}}_{nkI} (\eta,\phi)= \psi_{nk}^{I}(\eta )
\varphi^{\pi_{\mbox{\scriptsize c}}}_{k}(\phi),
\label{wvib}
\end{eqnarray}
where the radial part
\begin{equation}
\psi^I_{nk}(\eta )=\sqrt{\frac {2c\Gamma(n+1)}{\Gamma(n+2s+1)}}
e^{-c\eta^2/2}(c\eta^{2})^sL^{2s}_n(c\eta^2)\
\label{psieta1}
\end{equation}
involves generalized Laguerre polynomials in the variable $\eta$  with $c=\sqrt{BC}/\hbar$
and $s=(1/2)\sqrt{k^2+bX(I,K,\pi a)}$. The angular part in the variable $\phi$ appears with a
positive or negative parity, $\pi_{\mbox{\scriptsize c}}=(\pm )$, of the even-even core as
follows
\begin{eqnarray}
\varphi_{k}^{\pi_{\mbox{\scriptsize c}}=(+)}(\phi)&=& \sqrt{2/\pi}\cos (k\phi ) \ , \qquad k=1, 3, 5,
...\ ,\label{parplus} \\
\varphi_{k}^{\pi_{\mbox{\scriptsize c}}=(-)}(\phi)&=& \sqrt{2/\pi}\sin (k\phi ) \ , \qquad k=2, 4, 6,
...\ . \label{parminus}
\end{eqnarray}

The energy spectrum is determined in (\ref{enspect}) by the quantum numbers $n$ and $k$.
The parity-doublet structure is imposed by the condition $\pi=\pi_{\mbox{\scriptsize
c}}\cdot\pi_{\mbox{\scriptsize sp}}$, where $\pi_{\mbox{\scriptsize sp}}$ is the parity of the
odd-nucleon (single-particle) state. Then for a given state belonging to a parity-doublet the
core parity is determined as $\pi_{\mbox{\scriptsize c}}=\pi\cdot\pi_{\mbox{\scriptsize sp}}$.
The total core plus particle wave function for a state with an angular momentum $I^{\pi}$
belonging to a parity-doublet sequence in odd-even nuclei is given by \cite{b2b3odd}
\begin{eqnarray}
\Psi^{\pi ,\pi_{\mbox{\scriptsize sp}}}_{nkIMK}(\eta ,\phi)&=&
\sqrt{\frac{2I+1}{16\pi^2}}\Phi^{\pi\cdot \pi_{\mbox{\scriptsize sp}}}_{nkI} (\eta,\phi)
\left[ D^{I}_{M\, K}(\theta )\mathcal{F}_K \right. \nonumber \\
&+& \left. \pi\cdot\pi_{\mbox{\scriptsize sp}}(-1)^{I+K}
D^{I}_{M\,-K}(\theta )\mathcal{F}_{-K}\right ]\ ,
\label{spwf}
\end{eqnarray}
where $D^{I}_{M\, K}(\theta )$ is the rotation Wigner function and $\mathcal{F}_K$ is the
wave function of the odd-nucleon state.

In the present work it is supposed that the odd-nucleon parity
$\pi_{\mbox{\scriptsize sp}}$ is a good quantum number, although in the more
general treatment of the single-particle motion in the octupole deformed
(reflection-asymmetric) mean-field (potential) the single nucleon state may
appear with a mixed parity \cite{Leand82}-\cite{LC88}, \cite{MDSS10}. Then the
parity-doublet is determined by a given $n$ and a pair of odd and even
$k$-values, $k^{(\pi_{\mbox{\scriptsize c}}=+)}_{n}$ and
$k^{(\pi_{\mbox{\scriptsize c}}=-)}_{n}$ ($k^{(+)}_{n}<k^{(-)}_{n}$),
respectively. The $k$-values are determined so that $k=k^{(+)}_{n}$ for
$I^{\pi=\pi_{\mbox{\scriptsize sp}}}$ and $k=k^{(-)}_{n}$ for
$I^{\pi=-\pi_{\mbox{\scriptsize sp}}}$.  The difference between $k^{(+)}_{n}$
and $k^{(-)}_{n}$ generates  in (\ref{enspect})  an energy splitting of the
parity-doublet. That is why the term ``quasi parity-doublet'' was involved.
According to the rule above the states having the same parity as the ground or
bandhead state appear lower in energy and are characterized by the odd
$k^{(+)}_{n}$ number, while the opposite-parity states are shifted up and are
labeled by an even (and larger) $k^{(-)}_{n}$ number. The yrast doublet with
$n=0$ is formed on the top of the ground state whose parity is
$\pi_{\mbox{\scriptsize sp}}^{(n=0)}$. The non-yrast doublets with $n=1,2,...$
are coupled to excited s.p. or quasi-particle (q.p.) states (if the pairing
correlations are taken into account) whose parities $\pi_{\mbox{\scriptsize
sp}}^{(n)}$ determine the respective quasi-doublet structures according to the
rule above. Also, the index $n$ labels the different intrinsic configurations
to which the non-yrast doublets are coupled. The above model mechanism of the
forming of parity-doublet structures takes into account the possibility for a
switch, at certain higher angular momentum, of the s.p. state to which the
doublet states are coupled to a state with an opposite parity,
$\pi_{\mbox{\scriptsize sp}}\rightarrow -\pi_{\mbox{\scriptsize sp}}$. As
suggested in \cite{b2b3odd} the switch can be explained as the effect of an
alignment process in the core \cite{JMS05}. In the present model this leads to
a respective inversion of the parity $\pi_{\mbox{\scriptsize c}}$ of the
vibration state at the given angular momentum and to a subsequent inversion of
the mutual disposition of the parity-doublet counterparts, up-down
$\leftrightarrow$ down-up. Such a situation is indeed observed in few cases and
described by the model (see next section).

In the original application of the model to the yrast octupole spectra it was considered that for
a given $n$ the nucleus always takes the lowest quadrupole-octupole vibration states with
angular phonon numbers $k=1$ or $2$ depending on the parity \cite{b2b3mod,b2b3odd}.
The generalization of the model description by including non-yrast alternating-parity spectra
of even-even nuclei showed that a better agreement with the experimental data can be
obtained if higher $k^{(\pm)}_{n}$-values are also allowed \cite{MDSSL12}. It was seen that
the difference $\Delta k_{n}=k^{(-)}_{n}-k^{(+)}_{n}$ reasonably specifies the mutual
disposition of the opposite-parity sequences in the spectrum. Therefore, it can be expected
that the same meaning of the higher $k$-values will be valid for the parity-shift in the
quasi-doublet bands of odd-mass nuclei. On the other hand, as recognized in
\cite{MDSSL12}, in this case one has jumps of the quantum numbers $k_{n}^{(+)}$  and
$k_{n}^{(-)}$ over several low-lying angular-phonon excitations within the set of levels
characterized by given radial-phonon number $n$. Presently this is only justified by the
meaning of $\Delta k_{n}$ as a characteristics of the mutual displacement of the
opposite-parity sequences. The search for a deeper meaning and more sophisticated
correlation between the quadrupole and octupole modes capable to compensate or explain
these jumps is still an open issue. In this work we apply the original concept for the lowest
$k^{(\pm)}_{n}=1$ or $2$ phonon numbers in the model description of yrast and non-yrast
quasi parity-doublet spectra of odd-$A$ nuclei. At the same time in the end of the next section
we discuss the difference between the obtained results and the result of calculations
performed for the same nuclei by allowing higher $k_{n}^{(-)}$ values for the upper shifted
doublet counterparts and fixed $k^{(+)}_{n}=1$ value for the counterparts whose parity
coincides with the bandhead parity. For this reason we keep the formalism in its general form
capable of treating unrestricted values of angular-phonon numbers.

By using the wave functions (\ref{spwf}) one can calculate B(E$\lambda $) transition
probabilities ($\lambda =1,2,3$) in the yrast and non-yrast quasi-doublet spectra. The relevant
formalism was originally developed in \cite{b2b3mod,b2b3odd} and further extended to the
non-yrast states of even-even nuclei \cite{MDSSL12}. In this work we apply the formalism
developed in \cite{MDSSL12}. Due to the imposed axial symmetry the B(E$\lambda $)
probabilities are non-zero only between states with the same $K$-values. Also, here the
odd-nucleon wave function, $\mathcal{F}_K$ in (\ref{spwf}), is not considered explicitly,
while its parity is taken into account through the above explained parity-coupling scheme.  As
a result in this work we consider transition probabilities between states belonging to the same
yrast or non-yrast quasi parity-doublet and, therefore, coupled to the same s.p. state (doublet
bandhead). Then the CQOM reduced electric transition probabilities B($E\lambda$) with
multipolarity $\lambda =1,2,3$ between the initial (i) and final (f) doublet states have the form
\begin{eqnarray}
&B(E\lambda;n_i k_i I_{i} K_i\rightarrow n_f k_f I_{f} K_f) \hspace{3.8cm} \nonumber \\
&\hspace{-0.6cm} =\frac{2\lambda +1}{4\pi (4-3\delta_{\lambda,1})}
\langle I_iK_i\lambda 0|I_fK_f\rangle^2
R_{\lambda}^{2}(n_i k_i I_{i}\rightarrow n_f k_f I_{f}),
\label{BELA}
\end{eqnarray}
where $K_{i}=K_{f}$ corresponds to the same s.p. state. The factors $R_{\lambda}$ involve
integrals over the radial $\eta$ and the angular $\phi$ variables and can be written in the
following compact form
\begin{eqnarray}
&R_{\lambda}(n_i k_i I_{i}\rightarrow n_f k_f I_{f})\hspace{5.3cm}  \nonumber \\
&=M_{\lambda}p^{a_{\lambda}}q^{b_{\lambda}}
S_{l_{\lambda}}(n_i k_i I_i;n_f k_f  I_f)I_{\lambda}^{\pi_i,\pi_f}(k_i,k_f) ,
\label{rlam}
\end{eqnarray}
with the exponents $a_{\lambda}=1,1,0$, $b_{\lambda}=1,0,1$ and indexes
$l_{\lambda}=2,1,1$ for $\lambda =1,2,3$\,, respectively. The integrals over $\eta$ are
\begin{eqnarray}
&S_{l}(n_i k_i I_i;n_f k_f  I_f) \hspace{5.8cm}  \nonumber \\
&=\int_0^{\infty} d\eta \psi_{n_f k_f}^{I_f}(\eta)\eta^{l+1}
\psi_{n_i k_i}^{I_i}(\eta), \ \ \ l=1,2
\label{si}
\end{eqnarray}
while the integrals over $\phi$ read
\begin{eqnarray}
\hspace{-0.5cm}
I_{\lambda}^{\pi_i,\pi_f}(k_i,k_f)&=&\frac{2}{\pi}\int_{-\frac{\pi}{2}}^{\frac{\pi}{2}}
A_{\lambda}(\phi)\varphi^{\pi_f \cdot\pi_{\mbox{\scriptsize sp}}}_{k_f}(\phi)
\varphi^{\pi_i \cdot\pi_{\mbox{\scriptsize sp}}}_{k_i}(\phi)d\phi , \
\label{Ilam}
\end{eqnarray}
where the factors $A_{\lambda }(\phi)$ represent series expansions defined in Eqs.~(28)-(30)
of \cite{MDSSL12}. Analytic expressions for the  $\eta$-integrals  (\ref{si}) and explicit
expressions for the $\phi$-integrals are given in Appendices B and C of \cite{MDSSL12},
respectively. The quantities $R_{\lambda}$ depend on the multipole charge factors
$M_{\lambda}$, as determined in Eqs. (22) and (23) of \cite{MDSSL12}. The parameters $p$
and $q$ are defined in Eq.~(\ref{pqd}) above. However as shown in \cite{MDSSL12} they
are not independent but one has $q=p/\sqrt{2p^{2}-1}$, so that only $p$ is kept as an
adjustable parameter.  Also, the integrals (\ref{si}) depend on the parameter $c$ which enters
the radial wave functions (\ref{psieta1}). In addition, an effective charge
$e^{1}_{\mbox{\scriptsize eff}}$ is introduced to determine the correct scale of the B(E1)
transition probabilities with respect to B(E2) \cite{MDSSL12}.

\begin{figure*}
\centering\centerline{\includegraphics[scale=0.55]{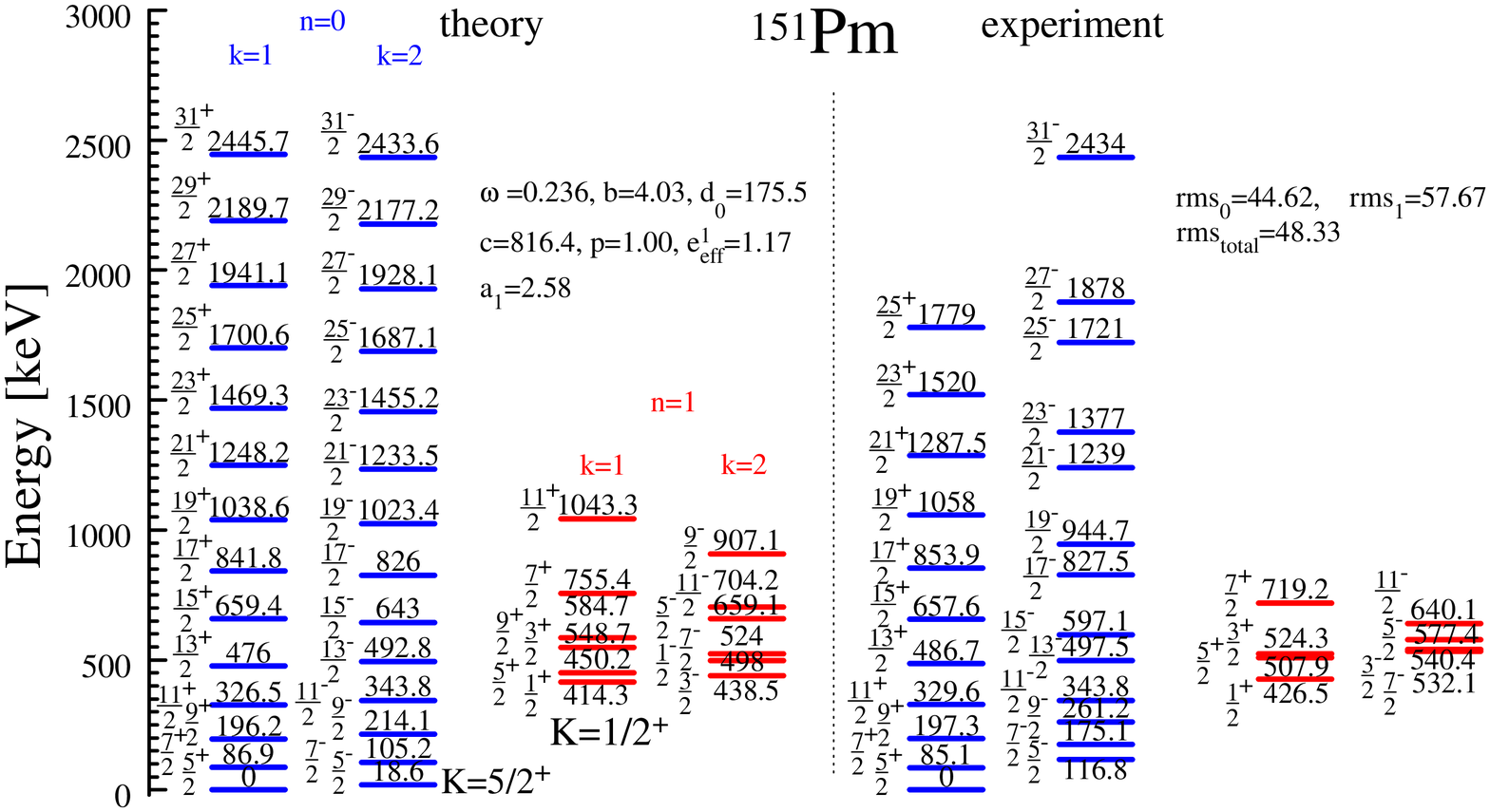}}
\centering\centerline{\includegraphics[scale=0.55]{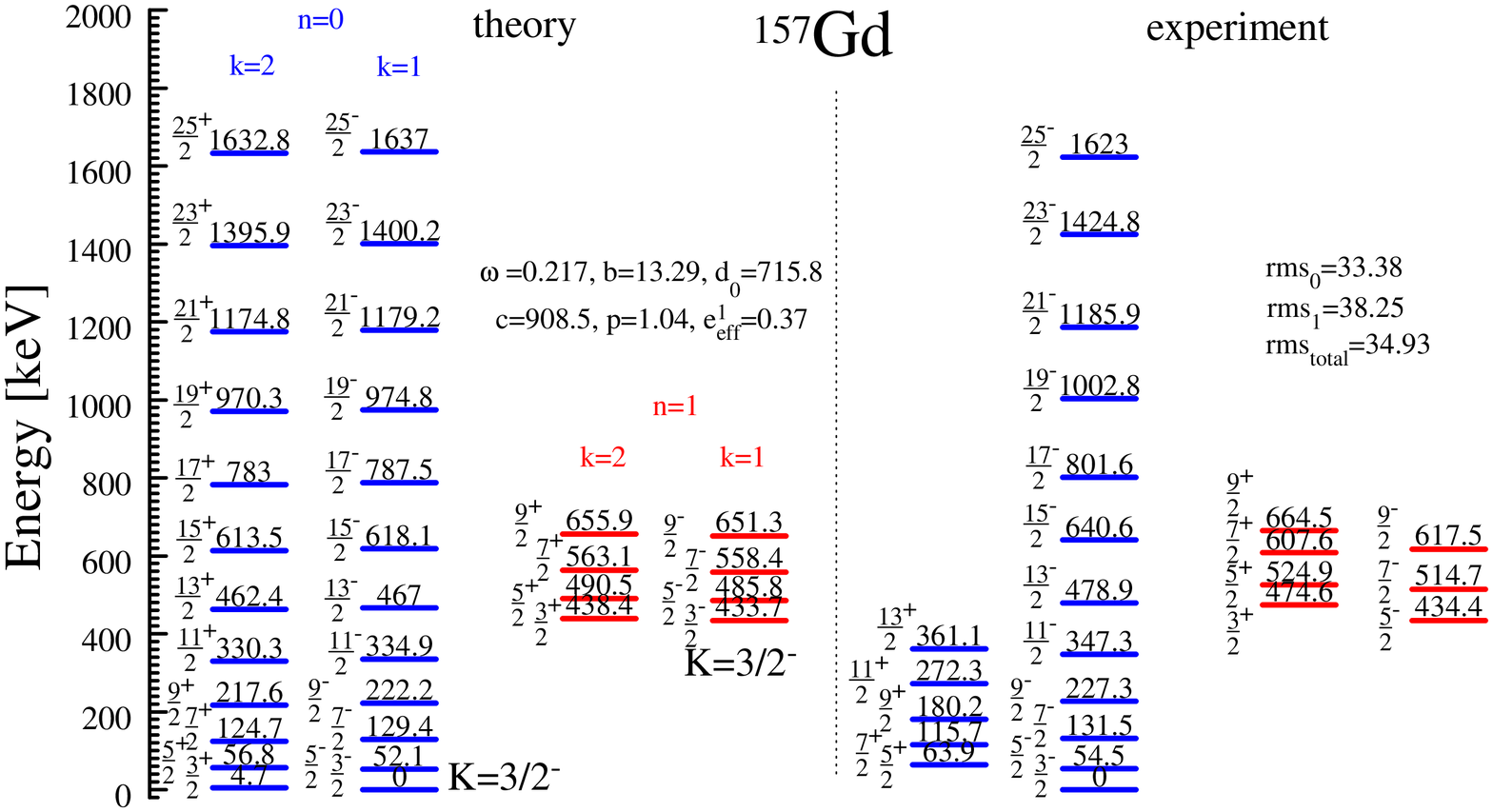}} \caption{Theoretical and
experimental quasi parity-doublet levels for $^{151}$Pm and $^{157}$Gd. The root mean
square (rms) deviations between the theoretical and experimental levels of the yrast band
(rms$_{0}$), the first non-yrast band (rms$_{1}$) and the total rms factor for both bands
(rms$_{\mbox{\scriptsize total}}$) are given in keV. Data from \cite{ensdf}. See the text for
details.} \label{f1}
\end{figure*}

The consideration of transition probabilities between states belonging to different
parity-doublets, and with $K_{i}\neq K_{f}$, can be implemented after obtaining
$\mathcal{F}_K$ within the reflection-asymmetric deformed shell model (DSM)
\cite{qocsmod}  as done in \cite{MDSS10} and by taking into account the Coriolis
$K$-mixing effect in the respective s.p. states as proposed in \cite{Mi13}. Such a task implies
a natural connection of the collective CQOM model to an approach in which the ground state,
the excited q.p. states, the respective decoupling factors (for $K=1/2$), as well as the Coriolis
mixing contributions are obtained microscopically.  Although the work in this direction is in
an essential progress \cite{Mi13}, here we consider the s.p. degree of freedom
phenomenologically. We consider that the knowledge of the pure collective CQOM
description of quadrupole-octupole spectra, especially the yrast and non-yrast quasi-doublet
levels, is a necessary step before attempting detailed implementation of the microscopic part
of the approach. Thus for a given doublet we take the s.p. parity and the $K$-value as
suggested by the experimental analysis or by microscopic calculations reported in the
literature. The quasi-doublet band-heads are obtained as different rotation-vibration modes
characterized by the CQOM oscillator quantum number $n=0,1,2,...$. In the cases of $K=1/2$
the decoupling factors $a_{n}$ for the respective doublets, labeled by $n$ and entering the
expression (\ref{Xdecoup}), are adjusted according to the experimental data. It should be
noted that these phenomenological decoupling factors are of special importance for
determining the physically reasonable deformation regions where DSM calculations have to
be performed after inserting the microscopic part in the CQOM. (See \cite{MDSS10} for
more details on this consideration.)

\begin{figure*}
\centering \centerline{\includegraphics[scale=0.55]{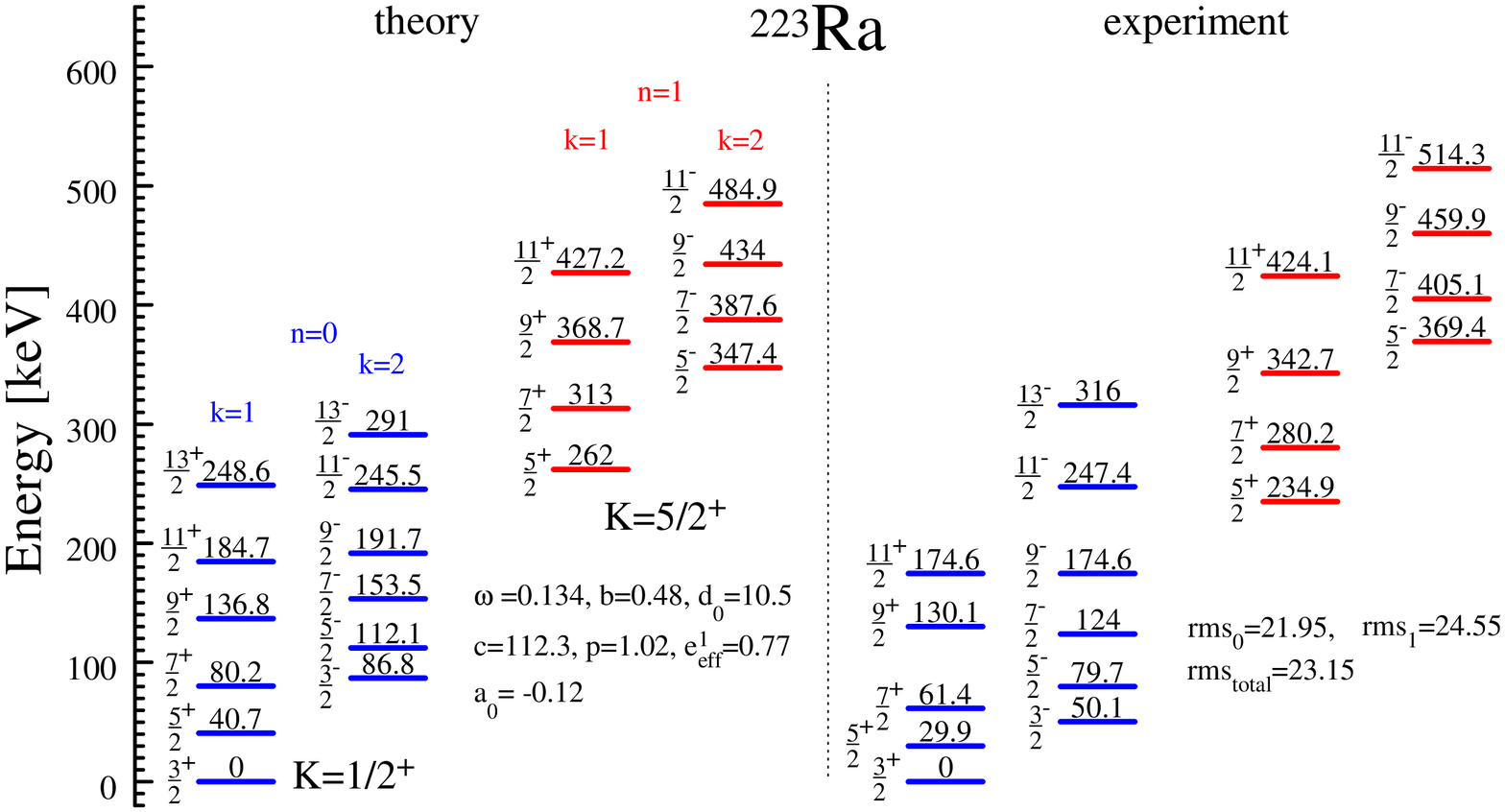}} \caption{The same as
Fig.~\ref{f1}, but for $^{223}$Ra.} \label{f2}
\end{figure*}

\section{Numerical results and discussion}
\label{sec:3}

\begin{figure*}
\centering \centerline{\includegraphics[scale=0.55]{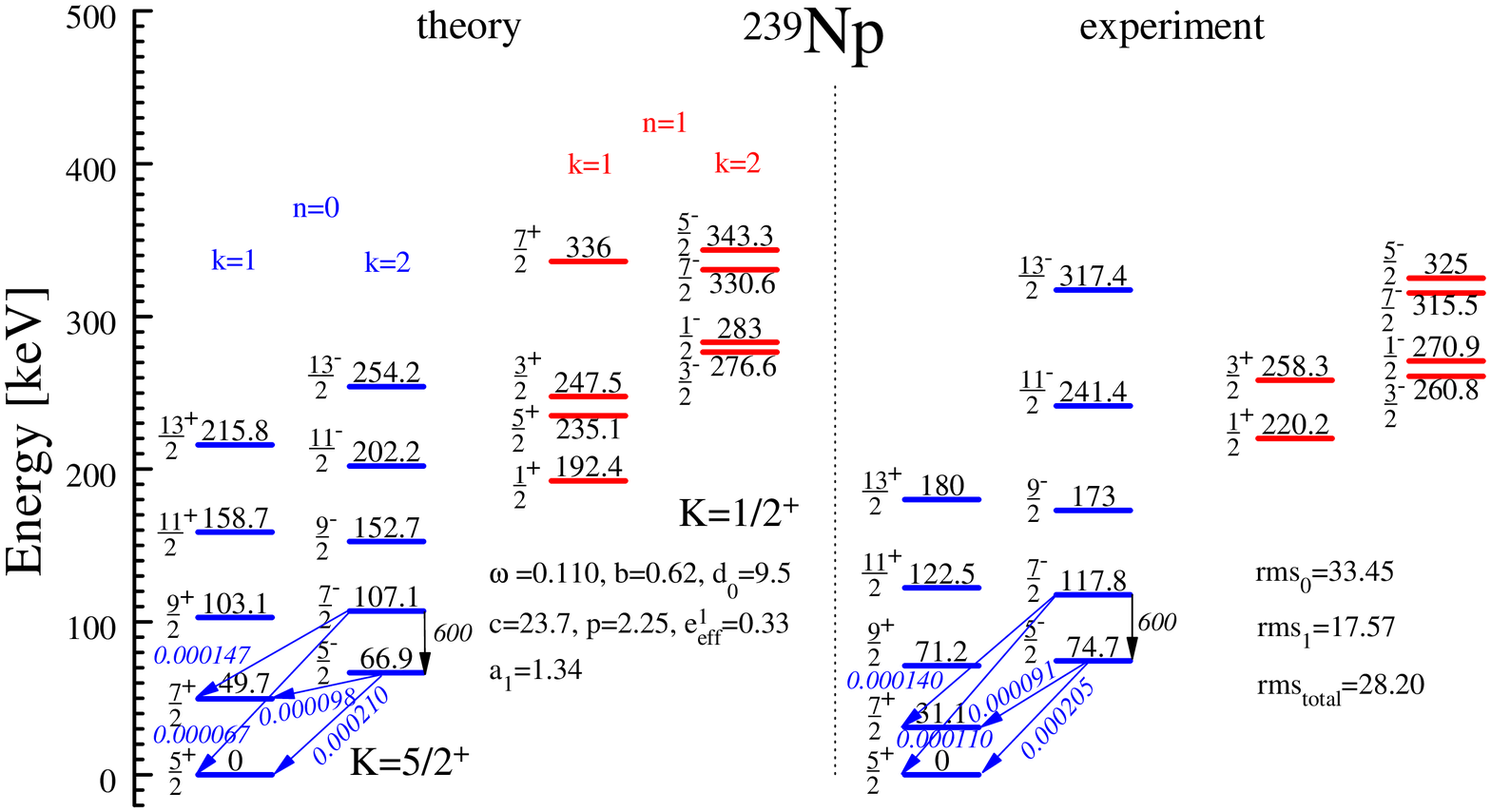}}
\centering\centerline{\includegraphics[scale=0.55]{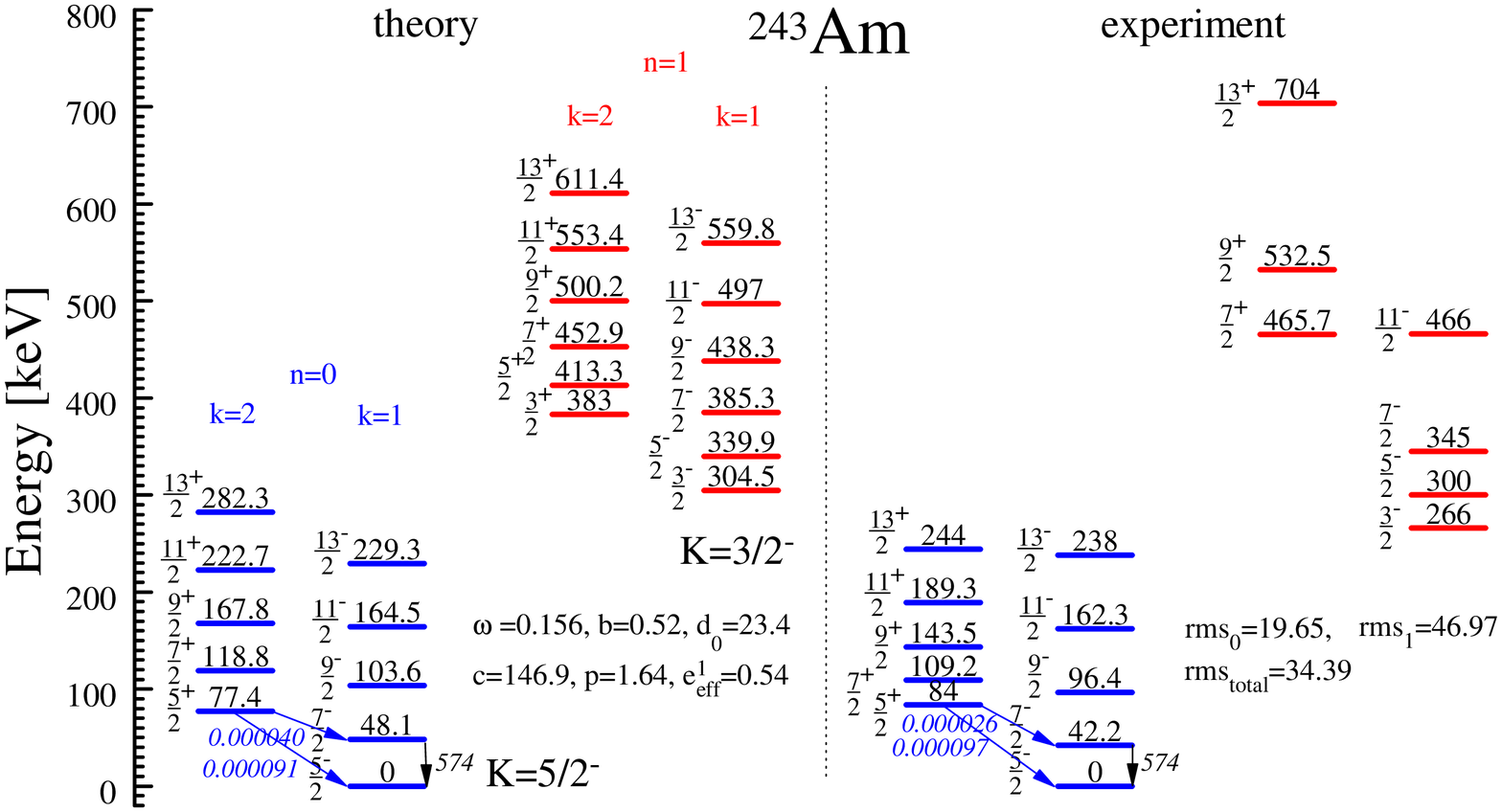}} \caption{The same as
Fig.~\ref{f1}, but for $^{239}$Np and $^{243}$Am. Theoretical and experimental B(E1) and
B(E2) transition probabilities are also given (data from \cite{nudat2}).} \label{f3}
\end{figure*}

In this section results of the CQOM model calculations for several odd-mass nuclei are
presented. The model energy levels are determined by Eq.~(\ref{enspect}) as
\begin{eqnarray}
\hspace{-0.7cm}\tilde{E}_{nk}(I,K,\pi a) =E_{nk}(I,K,\pi a)-E_{0k_{0}}
(I_{0},K_{0},\pi_{0} a_{0}),
\end{eqnarray}
where $I_{0}$,  $K_{0}$, $k_{0}=1$, $\pi_{0}$ and $a_{0}$ correspond to the
ground state. The model parameters $\omega$, $b$ and $d_{0}$ determine the
energy levels, while the parameters $c$, $p$ and the effective charge
$e^{1}_{\mbox{\scriptsize eff}}$ determine in addition the transition
probabilities as explained below Eq.~(\ref{Ilam}). These parameters are
adjusted by simultaneously taking into account experimental data on the energy
bands \cite{ensdf} and the available B(E1) and B(E2) transition probabilities
\cite{nudat2}. (Data on B(E3) probabilities are not available.) The theoretical
B(E1) and B(E2) values are calculated through Eq.~(\ref{BELA}). In the case(s)
of $K=1/2$ bandheads the decoupling factor(s) $a_{n}$ ($n=0$ for the yrast, and
$n=1$ for the first non-yrast doublets) is (are) also adjusted. Calculations
were performed for the nuclei $^{151}$Pm, $^{157}$Gd, $^{223}$Ra, $^{239}$Np
and $^{243}$Am. For each of them the yrast and one non-yrast quasi
parity-doublet are considered together with a number of experimentally observed
B(E1) and B(E2) transition probabilities.

The theoretical and experimental energy levels of the nuclei $^{151}$Pm,
$^{157}$Gd and $^{223}$Ra are compared in Figs.~\ref{f1} and \ref{f2}. In
Fig.~\ref{f3} both the energy levels and the transition probabilities for the
nuclei $^{239}$Np and $^{243}$Am are given.  The obtained parameter values are
given in the figures. Also, the root mean square (rms) deviations between the
theoretical and experimental levels of the yrast band (rms$_{0}$), the first
non-yrast band (rms$_{1}$) and the total rms factors for both bands
(rms$_{\mbox{\scriptsize total}}$) are given in Figs.~\ref{f1}-\ref{f3}. The
theoretical and experimental B(E1) and B(E2) transition probabilities for all
considered nuclei are compared in Table 1. For each nucleus the model
classification scheme accommodates one excited non-yrast quasi parity-doublet
band. In all nuclei the theoretical energy sequences reproduce the structure of
the experimentally observed yrast and non-yrast bands. This makes it meaningful
to predict a few more states in some of the considered bands in order to show
how the respective doublet-structures could develop with the angular momentum.
More specifically, in Figs.~\ref{f1}-\ref{f3} we add a number of predicted
states to one of the parity-counterpart sequences in order to get an equal
number of positive- and negative-parity levels in the theoretically obtained
quasi-doublets.

The quasi-doublet sequences in the nuclei $^{151}$Pm and $^{157}$Gd given in
Fig.~\ref{f1} have similar overall structures. In $^{151}$Pm we have a
relatively well developed yrast doublet built on a $K=5/2^{+}$ ground state and
a shorter non-yrast doublet considered to be coupled to a $K=1/2^{+}$ s.p.
state. In the latter a strong Coriolis decoupling effect is observed, such that
some higher-spin states appear at lower energy. This spin-interchange effect is
partly reproduced in the model description through a relatively large
decoupling-parameter value $a_{1}=2.58$. At the same time in the experimental
yrast sequence we indicate an inversion of the up-down shift between the
opposite-parity counterparts at $I^{\pi}=15/2^{\pm}$. This effect corresponds,
according to the model assumption explained in the previous section, to an
inversion of the s.p. parity and imposes an interchange of the $k^{(+)}_{0}=1$
and $k^{(-)}_{0}=2$ values in the model spectrum for $I^{\pi}\geq 15/2^{\pm}$.
As a result the observed shift of the positive-parity states above their
negative-parity counterparts is reproduced.

In $^{157}$Gd both quasi-doublets are built on negative-parity s.p. states with
$K=3/2^{-}$ which leads to the shift of the positive-parity counterparts in the
lower part of the spectrum up in energy compared to the negative-parity ones.
This is described according to the model scheme by assigning $k^{(-)}_{n}=2$ to
the positive-parity states and $k^{(+)}_{n}=1$ ($n=0,1$) to the negative-parity
ones. Similarly to $^{151}$Pm, in the yrast sequence of $^{157}$Gd the two
$k^{(\pm)}_{0}$ values are interchanged at $I^{\pi}=7/2^{\pm}$. We remark that
for both nuclei in Fig.~\ref{f1} the two $k$-values given above the theoretical
yrast-band sequences correspond to the respective positive- or negative-parity
counterparts before the interchange.

In $^{223}$Ra,  given in Fig.~\ref{f2}, the yrast band is described by imposing
$K=1/2^{+}$ instead of the experimentally suggested $K=3/2^{+}$. This is
motivated by the previously indicated staggering behavior of the doublet
splitting which is considered a manifestation of a Coriolis decoupling effect
in that band \cite{b2b3odd}. We remark that the decoupling factor obtains the
same value $a_{0}=-0.12$ as in the model description obtained in \cite{b2b3odd}
for the yrast band only. (In Fig. 3(b) of \cite{b2b3odd} this value was
misprinted without the correct sign $(-)$.) The considered non-yrast
quasi-doublet is built on a $K=5/2^{+}$ bandhead state. We see that the overall
disposition of this doublet with respect to the yrast band is reasonably
reproduced. The parity splitting in the yrast sequence is a bit overestimated
by the theory, while that in the non-yrast band is slightly underestimated.
Nevertheless, the root mean square (rms) deviations between the theoretical and
experimental levels of the different bands as well as the total rms factor are
obtained in quite reasonable limits between 20 and 25 keV (see Fig.~\ref{f2}).

The description of $^{239}$Np and $^{243}$Am given in Fig. 3 illustrates the applicability of
the model scheme in the region of heavier nuclei. Here we can remark that the strong Coriolis
effect in the non-yrast band of $^{239}$Np is adequately described with a decoupling factor
$a_{1}=1.34$. As a result the situating of the $3/2^{-}$ state below the $1/2^{-}$ state and the
$7/2^{-}$ state below the $5/2^{-}$ state in this band is well reproduced. At the same time the
model predicts that in the positive-parity sequence of the non-yrast band the $5/2^{+}$ state
will be placed below the $3/2^{+}$ state.

The theoretical values of the B(E1) and B(E2) transition probabilities, given
in Table~I, show an overall good agreement between theory and experiment.  This
makes it meaningful to predict a few more transition-probability values,
especially between the lowest opposite-parity states in the non-yrast doublets
(E1 transitions) as well as between states within sequences with the same
parity (E2 transitions). Such predictions are given for all considered nuclei.
It should be noted that data on transition probabilities in the non-yrast parts
of the considered spectra are not available. In addition, we should remind that
transitions between states belonging to different quasi parity-doublets are not
included in the present consideration. For a number of yrast-band transitions
larger discrepancies between theory and experiment are observed. In some cases,
such as the E1 transitions in $^{223}$Ra, this can be explained with the more
complicated structure due to the presence of the Coriolis interaction. However,
we remark on the good model reproduction of the experimental B(E1) transition
values in $^{151}$Pm,  $^{239}$Np and $^{243}$Am. In the nuclei $^{239}$Np and
$^{243}$Am for which only one B(E2) value is described we have an exact
agreement between the theory and experiment due to the exact determination of
the parameters $c$ and $p$ in the fitting procedure.

\begin{table}

\caption{Theoretical and experimental values of B(E1) and B(E2) transition probabilities in
Weisskopf units (W.u.) for quasi parity-doublet spectra of several odd-mass nuclei.
Notations: $I_{n_{i}}^{\pi_{i}}\rightarrow I_{n_{f}}^{\pi_{f}}$ with $n_{i}$ and $n_{f}$
denoting the doublet ($n=0,1$). The theoretical values are calculated in Eq.~(\ref{BELA}).
The data are taken from \cite{nudat2}. The uncertainties (in parentheses) refer to the last
significant digits in the experimental data.}

\begin{center}

{\small

\begin{tabular}{cccc}

\hline Mult& Transition & Th [W.u.] & Exp [W.u.]\\
\hline \hline

\multicolumn{4}{c}{$^{151}$Pm}\\

E1&$5/2^{-}_{0}$ $\rightarrow$ $5/2^{+}_{0}$ & 0.0011&0.0014 (2)\\

E1&$7/2^{-}_{0}$ $\rightarrow$ $5/2^{+}_{0}$ & $3.3\times 10^{-4}  $ &$>1.2  \times 10^{-4}$\\

E1&$7/2^{-}_{0}$ $\rightarrow$ $7/2^{+}_{0}$ & $6.3 \times 10^{-4} $ &$>2.1  \times 10^{-4}$\\

E2&$7/2^{-}_{0}$ $\rightarrow$ $5/2^{-}_{0}$ & 8 &$>8$\\

E2&$7/2^{+}_{0}$ $\rightarrow$ $5/2^{+}_{0}$ & 85 &\\

E1&$3/2^{-}_{1}$ $\rightarrow$ $1/2^{+}_{1}$ & $6.3  \times 10^{-4}$& \\

E1&$5/2^{-}_{1}$ $\rightarrow$ $3/2^{+}_{1}$ & $7.7  \times 10^{-4}$& \\

E2&$3/2^{+}_{1}$ $\rightarrow$ $1/2^{+}_{1}$ & 51&\\

\multicolumn{4}{c}{$^{157}$Gd}\\

E1&$5/2^{+}_{0}$ $\rightarrow$ $3/2^{-}_{0}$ & $46.1  \times 10^{-7}$  &$4.6 \times 10^{-7}$  (8)\\

E1&$5/2^{+}_{0}$ $\rightarrow$ $5/2^{-}_{0}$ & $ 4.47 \times 10^{-6}$  &$10.3 \times 10^{-6}$ (22)\\

E2&$5/2^{-}_{0}$ $\rightarrow$ $3/2^{-}_{0}$ & 311 &293\\

E2&$7/2^{-}_{0}$ $\rightarrow$ $3/2^{-}_{0}$ & 130 &119\\

E2&$7/2^{-}_{0}$ $\rightarrow$ $5/2^{-}_{0}$ & 195 &230\\

E2&$7/2^{+}_{0}$ $\rightarrow$ $5/2^{+}_{0}$ & 18 &\\

E1&$5/2^{+}_{1}$ $\rightarrow$ $5/2^{-}_{1}$ & $ 4.7  \times 10^{-6}$&\\

E1&$7/2^{+}_{1}$ $\rightarrow$ $5/2^{-}_{1}$ & $ 6.6  \times 10^{-6}$&\\

E2&$5/2^{+}_{1}$ $\rightarrow$ $3/2^{+}_{1}$ & 30 &\\

E2&$7/2^{-}_{1}$ $\rightarrow$ $5/2^{-}_{1}$ & 199 &\\

\multicolumn{4}{c}{$^{223}$Ra}\\

E1&$3/2^{-}_{0}$ $\rightarrow$ $3/2^{+}_{0}$ & $4 \times 10^{-5}$ &$119\times 10^{-5}$ (16)\\

E1&$3/2^{-}_{0}$ $\rightarrow$ $5/2^{+}_{0}$ & $4.2 \times 10^{-4}$ &$5.0\times 10^{-4}$ (9)\\

E1&$7/2^{-}_{0}$ $\rightarrow$ $5/2^{+}_{0}$ & $3.30  \times 10^{-4}$&$0.79\times 10^{-4}$ (24)\\

E1&$7/2^{-}_{0}$ $\rightarrow$ $7/2^{+}_{0}$ & $1.4 \times 10^{-5}$ &$2.0\times 10^{-5}$ (22,-5)\\

E2&$7/2^{-}_{0}$ $\rightarrow$ $3/2^{-}_{0}$ & 17 & 10 (6)\\

E2&$7/2^{+}_{0}$ $\rightarrow$ $5/2^{+}_{0}$ & 18 & 70\\

E2&$7/2^{+}_{0}$ $\rightarrow$ $3/2^{+}_{0}$ & 148 & 44\\

E2&$11/2^{+}_{0}$ $\rightarrow$ $7/2^{+}_{0}$ & 210 & 280 (12)\\

E1&$5/2^{-}_{1}$ $\rightarrow$ $5/2^{+}_{1}$ & $1.1 \times 10^{-3}$&\\

E1&$5/2^{-}_{1}$ $\rightarrow$ $7/2^{+}_{1}$ & $4.9 \times 10^{-4}$&\\

E1&$7/2^{-}_{1}$ $\rightarrow$ $7/2^{+}_{1}$ & $6.9 \times 10^{-4}$&\\

E2&$7/2^{+}_{1}$ $\rightarrow$ $5/2^{+}_{1}$ & 298 & \\

E2&$7/2^{-}_{1}$ $\rightarrow$ $5/2^{-}_{1}$ & 32 & \\

\multicolumn{4}{c}{$^{239}$Np}\\

E1&$5/2^{-}_{0}$ $\rightarrow$ $5/2^{+}_{0}$ & $2.10\times 10^{-4}$ & $2.05\times 10^{-4}$ (11)\\

E1&$5/2^{-}_{0}$ $\rightarrow$ $7/2^{+}_{0}$ & $9.8 \times 10^{-5}$ & $9.1 \times 10^{-5}$ (4)\\

E1&$7/2^{-}_{0}$ $\rightarrow$ $5/2^{+}_{0}$ & $7   \times 10^{-5}$ & $\geq 11  \times 10^{-5}$\\

E1&$7/2^{-}_{0}$ $\rightarrow$ $7/2^{+}_{0}$ & $1.5 \times 10^{-4}$ & $\geq 1.4 \times 10^{-4}$\\

E2&$7/2^{-}_{0}$ $\rightarrow$ $5/2^{-}_{0}$ & 600  &$\geq 600$\\

E2&$9/2^{-}_{0}$ $\rightarrow$ $7/2^{-}_{0}$ & 562  &\\

E1&$1/2^{-}_{1}$ $\rightarrow$ $1/2^{+}_{1}$ & $2.1 \times 10^{-4}$ &\\

E1&$3/2^{-}_{1}$ $\rightarrow$ $1/2^{+}_{1}$ & $2.1 \times 10^{-4}$ &\\

E2&$1/2^{-}_{1}$ $\rightarrow$ $3/2^{-}_{1}$ & 915  &\\

\multicolumn{4}{c}{$^{243}$Am}\\

E1&$5/2^{+}_{0}$ $\rightarrow$ $5/2^{-}_{0}$ & $9.1 \times 10^{-5}$ & $9.7 \times 10^{-5}$(4)\\

E1&$5/2^{+}_{0}$ $\rightarrow$ $7/2^{-}_{0}$ & $4 \times 10^{-5}$   & $2.6 \times 10^{-5}$(3)\\

E2&$7/2^{-}_{0}$ $\rightarrow$ $5/2^{-}_{0}$ & 574  & 574 (9)\\

E2&$7/2^{+}_{0}$ $\rightarrow$ $5/2^{+}_{0}$ & 61   &\\

E1&$3/2^{+}_{1}$ $\rightarrow$ $3/2^{-}_{1}$ & $1.7 \times 10^{-4}$  &\\

E1&$3/2^{+}_{1}$ $\rightarrow$ $5/2^{-}_{1}$ & $1.2 \times 10^{-4}$  &\\

E1&$5/2^{+}_{1}$ $\rightarrow$ $5/2^{-}_{1}$ & $7.7 \times 10^{-5}$  &\\

E1&$7/2^{+}_{1}$ $\rightarrow$ $7/2^{-}_{1}$ & $4.7 \times 10^{-5}$  &\\

E2&$5/2^{-}_{1}$ $\rightarrow$ $3/2^{-}_{1}$ & 767  &\\

E2&$5/2^{+}_{1}$ $\rightarrow$ $3/2^{+}_{1}$ & 79   &\\

\hline \hline

\end{tabular}
}
\end{center}

\end{table}

Here we can summarize that in all considered nuclei the rms deviations between
the theory and experiment obtained for the different sequences as well as the
total rms factors prove the good quality of the model description. We remark
that the total rms factors do not exceed 50 keV, with the lowest
rms$_{\mbox{\scriptsize total}}$=23 keV being obtained for $^{223}$Ra and the
higher one, 48 keV, observed in $^{151}$Pm. As mentioned in the previous
section it can be expected that considerably better descriptions with lower rms
factors may be expected if higher $k_{n}^{(-)}> 2$ values are allowed for the
upper quasi-doublet counterparts. We have performed such calculations for the
same data in the same nuclei. For each nucleus the calculations were performed
on a net over $k^{(-)}_{n}$ $(n=0,1)$ varying between $2$ and $20$ while
$k^{(+)}_{n}$ was kept equal to 1.  In this way the $k^{(-)}_{n}$ values which
provide the best overall description of energy levels and transition
probabilities were determined. As a result, essentially lower energy rms
factors were obtained in: $^{223}$Ra with ``favored'' $k_{0}^{(-)}= 4$ and
$k_{1}^{(-)}= 6$ values and rms$_{\mbox{\scriptsize total}}$=7 keV; $^{239}$Np
with $k_{0}^{(-)}= 4$ and $k_{1}^{(-)}= 2$ and rms$_{\mbox{\scriptsize
total}}$=15 keV; $^{243}$Am with $k_{0}^{(-)}= 4$ and $k_{1}^{(-)}= 6$ with
rms$_{\mbox{\scriptsize total}}$=18 keV.  It is seen that in these nuclei the
rms$_{\mbox{\scriptsize total}}$ values obtained at $k_{n}^{(-)}> 2$ are
smaller than the values (given in Figs. 2 and 3) obtained for $k_{n}^{(-)}= 2$
by a factor between 2 and 3. For the nuclei $^{151}$Pm with ``favored''
$k_{0}^{(-)}= 6$ and $k_{1}^{(-)}= 2$ values and rms$_{\mbox{\scriptsize
total}}$=37 keV, and $^{157}$Gd with $k_{0}^{(-)}= 2$ and $k_{1}^{(-)}= 6$ and
rms$_{\mbox{\scriptsize total}}$=29 keV the improvement of the description is
not too pronounced (see the rms values in Fig.~\ref{f1}).

The conclusion from the above mentioned calculations is that indeed the
involvement of larger numbers of angular phonons leads to better model
descriptions. Nevertheless the result illustrated in Figs.~\ref{f1}-\ref{f3}
shows  that the imposing of the lowest quadrupole-octupole vibration modes with
$k^{(+)}_{n}=1$ and $k^{(-)}_{n}=2$ still provides quite adequate
interpretation and classification of the yrast and first non-yrast
quasi-doublet energy sequences in the considered odd-mass nuclei. This means
that the CQOM model is capable of reproducing the specific spectroscopic
properties related to the simultaneous manifestation of quadrupole and octupole
degrees of freedom in these nuclei. The use of the model with larger
angular-phonon numbers can extend its applicability in wider regions of nuclei
and quasi-doublet spectra, including higher non-yrast bands, but needs a
detailed justification of the presence of lower model states for which
experimental data are not observed.  It should be noted that even on the
present level of phenomenology, the CQOM model description provides a useful
basis for further developments. As it was discussed in the end of
Sec.~\ref{sec:2} the knowledge of the decoupling factors as well as the model
mechanism for the forming of parity-doublet spectra in odd-mass nuclei can
guide the inclusion of microscopic calculations in the study.

\section{Conclusion}

\label{sec:4}

In conclusion, the present work reports an application of the collective model
of Coherent Quadrupole and Octupole Motion (CQOM) for the description of yrast
and non-yrast quasi-doublet spectra in odd-mass nuclei. The calculations are
performed by considering a zero number of radial quadrupole-octupole phonons in
the yrast band and one radial-phonon excitation in the first non-yrast band. In
both bands the lowest possible angular-phonon modes in the motion of the
even-even core are considered. The results obtained for a number of odd-mass
nuclei in the rare-earth and actinide regions illustrate the capability of the
model to reproduce the structure of  yrast and non-yrast energy levels together
with the attendant B(E1) and B(E2) transition probabilities. The test of the
model by allowing a higher number of angular phonons shows a better agreement
with the experimental data, but still needs a justification of the jumps over
the lower phonon numbers. At the same time the collective rotation-vibration
structure of the spectra and the observed Coriolis decoupling effects are
adequately taken into account. On this basis the present CQOM model
descriptions can serve as a starting point for the application of a deeper
collective and microscopic approach in the exploration of nuclear
quadrupole-octupole collectivity. Work in this direction is in progress.

\ \ \

\begin{acknowledgments}
This work is supported by DFG and by the Bulgarian National Science Fund
(Contract No. DID-02/16-17.12.2009).
\end{acknowledgments}

\end{document}